\documentstyle[vsolj01,graphicx,natbib]{article}

\begin{document}

\title{Remarkable yellow supergiant variable T\lowercase{mz}V429}

\author{Taichi Kato}
\author{Department of Astronomy, Faculty of Science, Kyoto University,
       Sakyou-ku, Kyoto 606-8502, Japan}
\email{e-mail: tkato@kusastro.kyoto-u.ac.jp}

\author{Kesao Takamizawa}
\author{Variable Star Observers League in Japan (VSOLJ), 65-1 Oohinata,
       Saku-machi, Nagano 384-0502, Japan}
\email{e-mail: k-takamizawa@nifty.ne.jp}

\begin{abstract}
   We discovered that GSC 6554.559 is a previously unknown variable star,
and named as TmzV429.  We noticed that TmzV429 is identified with the
IRAS-selected proto-planetary nebula (PPN), IRAS 08005-2356, which is
undergoing a vigorous mass-loss episode.  The analysis of photometric
data suggests that TmzV429 a short-period pulsator, resembling a
high-latitude yellow supergiant variable.  TmzV429 is considered to be
one of rare objects caught in the rapid course of PPN evolution, and
shows one of the most striking mass-loss features among variable stars
in the PPN stage.  Since its evolutionary time-scale is estimated to be
quite short ($\sim$150 yrs), future observations of pulsations of TmzV429
is expected to provide an excellent opportunity to study the stellar
evolution in real time.
\end{abstract}

\section{Introduction}

   Proto-planetary nebulae (PPNe) are objects in transition between
the AGB stage and planetary nebula (PN) stage in stellar evolution
(for a review, see \cite{hri97}).  PPNe are astrophysical objects not
only important in studying the mass-loss from post-AGB stars and the
formation of PNe, but also are considered to related to some of
enigmatic high-latitude luminous yellow variables, such as RV Tau stars
and UU Her stars (for a recent review, see \cite{hrilu97}).

   TmzV429 (=GSC 6554.559)\footnote{The permanent designation V510 Pup
has been given.} is a variable star discovered by Takamizawa
\citep{tak99}  The J2000.0 coordinates are 08\h 02\m 40\s.71,
$-$24\deg 04\arcm 42\farcs 4.  \citet{tak99} reported small amplitude
variations with a total photographic range of variability of 11.7--12.2.
\citet{tak99} originally suspected that this star is an semiregular
variable.  We discovered that this variable star, inconspicuous at the
time of the variability announcement, is identified with a conspicuously
mass-losing central star of a PPN, IRAS 08005-2356.  We describe in this
paper the analysis of our photometric data and the astrophysical
implications of the present identification with a rapidly evolving PPN.

\section{Observations}

   The photographic observations by Takamizawa (Saku All Sky Survey, SASS)
using 10-cm F/4.0 twin patrol cameras and T-Max400 120 emulsions.
The magnitudes were determined by comparison with non-red GSC stars,
whose zero-point offset from Tycho-2 $V$ magnitudes have been estimated
to be +0.5 mag.  This offset is confirmed by comparison with a single
point $V$-band CCD measurement (Kiyota, private communication).  Since
a constant offset does not affect the confirmation of the variability and
period analysis, we use the original measurements in the following
analysis.  The typical error of single estimates is $\sim$0.2 mag, which
will not affect the following discussion.

\begin{figure}
  \begin{center}
  \includegraphics[angle=0,width=8cm]{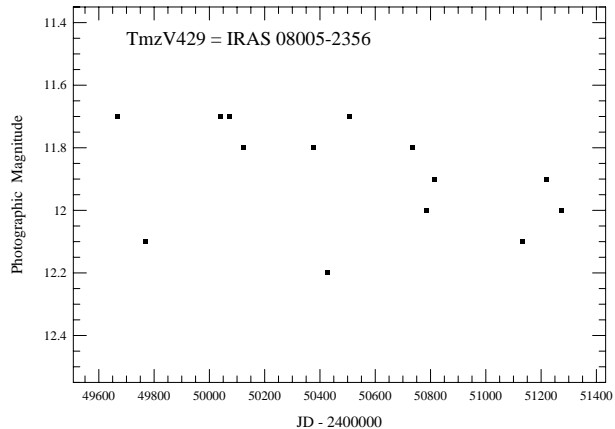}
  \caption{Light curve of TmzV429 drawn from the data in table \ref{tab:obs}.}
  \label{fig:lc}
  \end{center}
\end{figure}

   Table \ref{tab:obs} lists all observations of TmzV429 by this
observation.  Figure \ref{fig:lc} shows the overall light curve
drawn from these observations.

\begin{table}
\centering
\caption{Observations of TmzV429 by Takamizawa}\label{tab:obs}
\vskip 2mm
\begin{tabular}{cccccccc}
\hline\hline
JD$^a$ & mag$^b$ & JD$^a$ & mag$^b$ & JD$^a$ & mag$^b$ & JD$^a$ & mag$^b$ \\
\hline
49668.224 & 11.7 & 50125.040 & 11.8 & 50735.306 & 11.8 & 51218.037 & 11.9 \\
49769.035 & 12.1 & 50378.328 & 11.8 & 50786.225 & 12.0 & 51272.952 & 12.0 \\
50040.235 & 11.7 & 50426.155 & 12.2 & 50814.172 & 11.9 & & \\
50074.213 & 11.7 & 50506.995 & 11.7 & 51133.285 & 12.1 & & \\
\hline
 \multicolumn{8}{l}{$^{a}$ JD$-$2400000.} \\
 \multicolumn{8}{l}{$^{b}$ Photographic magnitude.  Close to $V$+0.5.} \\
\end{tabular}
\end{table}

\section{Discussion}

   We noticed that TmzV429 is identified with the PPN with a rapid
mass-loss, IRAS 08005-2356 \citep{sli91}.  The object is also identified
with an infrared source, MSX5C G242.3642+03.5822 \citep{ega99}.
The optical spectral classification by \citet{sli91} is a late F-supergiant
with prominent hydrogen emission lines.  Together with Takamizawa's
discovery of optical variability, the object seems to be classified as
an high luminosity yellow supergiant variable (SRD-type in the General
Catalogue of Variable Stars).

   \citet{hek99} also reported a possible brightening by a several tenths
of magnitude since 1986.  This possible variation seems to be more likely
attributed to shorter time-scale variations discovered by us.

\begin{figure}
  \begin{center}
  \includegraphics[angle=0,width=7cm]{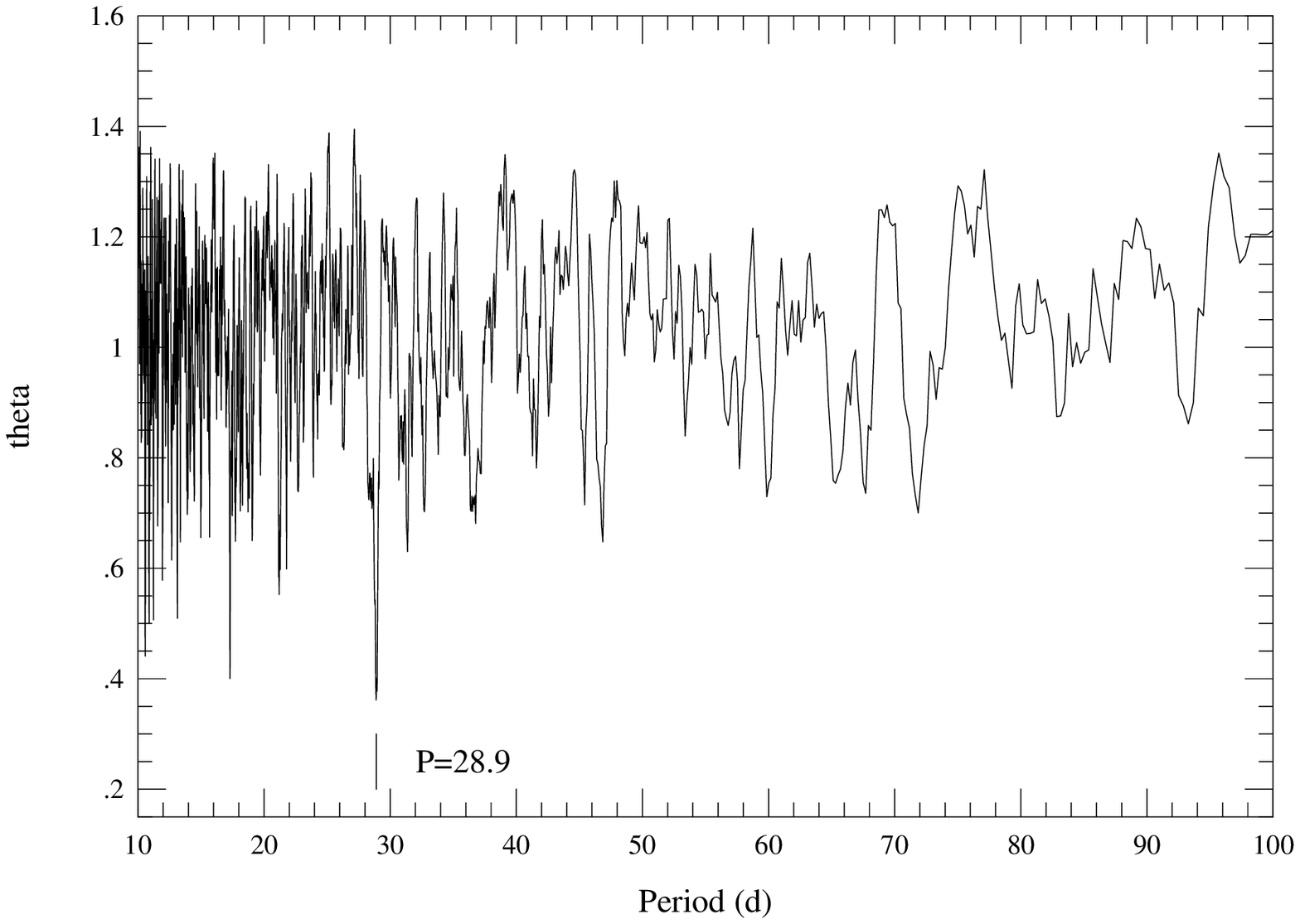}
  \includegraphics[angle=0,width=7cm]{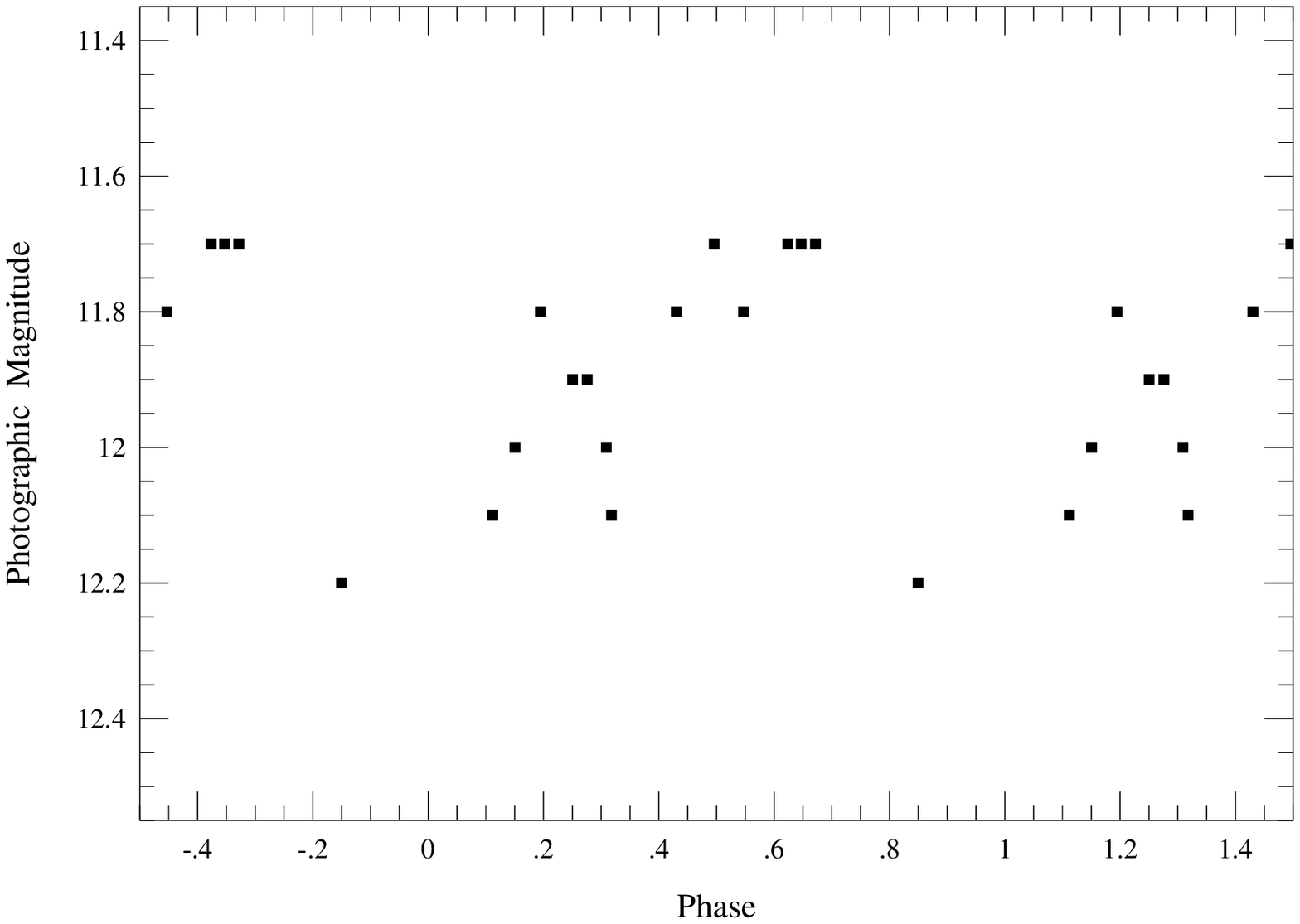}
  \caption{(Left) Period analysis TmzV429.  The most significant period of
  28.9 d is marked with a tick. (Right) Folded light curve of TmzV429.
  The phase zero is taken arbitrarily.}
  \label{fig:per}
  \end{center}
\end{figure}

   We analyzed the original discovery data by \citep{tak99} using
the Phase Dispersion Minimization (PDM) method \citep{ste78}.
The result of period analysis are shown in figure \ref{fig:per} (left
panel).  The period search was done for periods between 10 and 100 d.
The range was limited mainly due to the data sampling, but covers most
frequently met periods in low-mass, high luminosity, SRD-type variables.
The strongest period between 10 and 100 d is 28.9 d.  The period probably
needs be treated with caution, because the period is close to the lunar
month, and because of the possible intrinsic irregularity in such
a variable.  A rapid fading by 0.4 mag between JD 2450378 and 2450426,
however, supports the existence of short-period variation with a period
less than $\sim$100 d.

   The folded light curve by this period is shown in figure \ref{fig:per}
(right panel).  This result shows that the variability discovered by
Takamizawa can be expressed by oscillations with a single, relatively
short period.  Although the possibility of a longer period can not be
completely disregarded, the raw data (table \ref{tab:obs}) suggest a
short-period variation, rather than a period of hundreds of days to years.
\citet{sli91} reported some line features are similar to $\rho$ Cas.
The presently discovered variation, however, is not consistent with
variations with a much longer period ($\sim$300 d) as in massive $\rho$
Cas-like variables.  The star should be thereby regarded as a low-mass,
post-AGB pulsator (e.g. \cite{aik91}), which is consistent with the
evolutionary stage \citep{sli91} inferred from optical spectroscopy
and IRAS observations.

   Although the number of observations is still limited, and the present
analysis unavoidably suffers from a uncertainty, the present result
suggests the existence of a low-amplitude ($\sim$0.5 mag), relatively
short-period pulsations in TmzV429, which are analogous to variations
observed in some stellar components of other PPNe and in high galactic
luminous yellow variables, such as RV Tau stars and UU Her stars.
\citet{sli91} suggested that the evolutionary time scale of this object
is quite short ($\sim$150 yrs).  TmzV429 is thus one of rare objects
caught in the rapid course of PPN evolution, and shows one of the most
striking mass-loss features among variable stars in the PPN stage.
Pulsations in such stars are a sensitive indicator of the evolution
\citep{aik91}, future observations of pulsations of this object will
provide an excellent opportunity to study the {\it stellar evolution in
real time}.

\end{document}